 \documentclass[final,5p,times,twocolumn]{elsarticle}

\usepackage{amssymb}

\usepackage[table,xcdraw]{xcolor}
\usepackage{tcolorbox}
\usepackage{amsmath,amsfonts}
\usepackage{tipa}
\usepackage{url}
\usepackage{bm}
\usepackage{multirow}
\usepackage{hyperref}
\usepackage{graphicx}
\usepackage{adjustbox}
\usepackage{tabularx}
\usepackage{longtable}
\usepackage{bbding}
\usepackage{pifont}
\usepackage{wasysym}
\usepackage[utf8]{inputenc}
\usepackage{balance}

\journal{Pattern Recognition Letter}

\begin{document}

\begin{frontmatter}



\title{Normal-Anchored First-Order Model-Agnostic Meta-Learning based Whisper Fine-Tuning for Enhancing Fairness of
Cleft Lip and Palate Speech Recognition
}




\author[1]{\texorpdfstring{Susmita Bhattacharjee}}
\ead{sbhattacharjee@iitg.ac.in}

\author[2]{\texorpdfstring{Jagabandhu Mishra}}
\ead{jagabandhu.mishra@uef.fi}

\author[1]{H.S. Shekhawat}
\ead{h.s.shekhawat@iitg.ac.in}

\author[3,4]{Ravi Jasuja}
\ead{jasuja1@gmail.com}

\author[5]{S. R. Mahadeva Prasanna}
\ead{prasanna@iiitdwd.ac.in}


\affiliation[1]{organization={Department of Electronics and Electrical Engineering},
            addressline={Indian Institute of Technology Guwahati}, 
            city={Guwahati},
            postcode={781039}, 
            state={Assam},
            country={India}}

\affiliation[2]{organization={School of Computing},
            addressline={University of Eastern Finland}, 
            city={Joensuu},
            country={Finland}}         

\affiliation[3]{%
  organization={Brigham and Women's Hospital, Harvard Medical School},
  city={Boston},
  state={MA},
  country={United States}
}

\affiliation[4]{%
  organization={Function Promoting Therapies},
  city={Waltham},
  state={MA},
  country={United States}
}

\affiliation[5]{organization={Department of Electrical Engineering}, 
            addressline={Indian Institute of Information Technology Dharwad (IIIT Dharwad)}, 
            city={Dharwad},
            country={India}}
\begin{abstract}
Automatic speech recognition of CLP speech poses challenges because of variability of acoustics and articulation based on severity levels compared to typical speech. Consequently, this deteriorates the recognizing capability of ASR devices for Cleft Lip and Palate (CLP) speech. As a result finetuning of LLMs do not always yield good results. 
The reason behind the usage of First order model agnostic meta learning (FOMAML) is because of the low-resource and high-heterogeneity of the CLP speech data. As different CLP speech datasets differ greatly depending on the individual differences, and even the severity of the condition itself, a model trained with standard fine-tuning methods can have difficulties generalizing to other CLP groups. FOMAML presents a solution to this problem as it optimizes the model in order to adapt quickly - the inner loop simulates task-specific adaptation, while the outer loop encourages the adapted model to perform well on related but potentially more challenging speech conditions. Therefore, the method is well aligned with the goal of improving robustness and reliability in CLP ASR.
This work proposes Normal-Anchored First-Order Model-Agnostic Meta-Learning (NA-FOMAML), which applies fine-tuning of Whisper on CLP speech through the utilization of a first-order bilevel meta-learning framework. 
Under the proposed setup, normal speech was employed as the inner-loop input to ensure that a reliable source of adaptation was available, and CLP speech groups were incorporated in the outer loop to allow the model to converge on representations that retain their utility post-adaptation. This strategy was adopted with the aim of reducing the difference in performance between normal and pathological speech. In the inner loop, normal speech was utilized as the support set while different combinations of normal, mild, moderate and severe pathological speech were optimized at the outer loop level. The previous studies apply whisper fine-tuning of CLP speech dataset, but they suffer from high risks of overfitting. To address this problem, partial encoder fine-tuning is explored for both NMCPC and AIISH datasets using Frozen encoder followed by full encoder and selected encoder layers 0 to 5, 4 to 11, 6 to 11 and 8 to 11 of Whisper, coupled with decoder and projection head.
Four types of normal-anchored fine-tuning are examined in the study for two datasets NMCPC and AIISH. As observed in the experiment results, it can be seen that outer-loop training with only normal speech is insufficient for the recognition task.
In the case of NMCPC, the best performance is obtained by using full encoder layers with Normal to Normal+Mild+Moderate configuration having a WER of 4.40\%, 5.53\%, 16.14\% and 52.07\% for normal, mild, moderate and severe speech, respectively. In the case of AIISH dataset, the configuration that leads to optimal results is full encoder layers with Normal to Normal+Mild+Moderate+Severe having a WER of 2.48\%, 19.66\%, 14.05\% and 57.50\%. Besides WER analysis, an additional phoneme level error analysis is conducted using the transcription based phoneme category mapping method. It reveals that the errors in case of severe speech are very high and occur in fricatives, affricates, nasals, liquids, plosives and vowels. This concludes that NA-FOMAML is robust for cross-severity but needs further enhancement by employing severity-aware sampling, phoneme aware loss function and augmenting the pressure consonant and resonance related distortions for severe CLP speech.

\end{abstract}




\end{frontmatter}

\section{Introduction}

Speech disorder in an individual can arise either from craniofacial structural abnormalities as a birth defect resulting from deformation of craniofacial structure or from pathological conditions associated with a certain disease. Cleft Lip and Palate is one of such abnormalities which affects one in every 700 children worldwide. The reason for this speech defect is velopharyngeal dysfunction (VPD), oronasal fistula and mislearning~\cite{vallino2016evaluation, methofperceptualassessment, clpnatureandremediation}. Clinical interventions cannot fully fix the speech distortion which makes it difficult for CLP patients to communicate like a normal individual. Speech intelligibility in CLP patients is mainly degraded due to the presence of hypernasality and articulation error. This misarticulation due to the loss of oral pressure causes arrangement errors mainly for obstruents~\cite{clpeffectsonspeechandresonance, cpspeech, scp, Zajac2011ReliabilityAV, Whitehill2004SinglewordII}.  Severe nasality due to the large velopharyngeal gap results in the replacement of nasal consonants by the obstruents~\cite{evaluationand}, \cite{clpeffectsonspeechandresonance}.  This fairness aspect for pathological speech is not explored much for the existing ASR devices. As a result this type of speech becomes difficult to perceive not just by other normal listeners but also the publicly available ASR systems~\cite{fairnessmy}. One of such large-scale model released by OpenAI in 2022 is Whisper. It is a weakly-supervised encoder-decoder Transformer model trained on approximately 680,000 hours of multilingual audio data collected from the web. Whisper is an end-to-end sequence-to-sequence model capable of multitasking: transcription, translation to English, voice activity detection, language identification and timestamp prediction all within a single unified architecture~\cite{whisper}.

For our research, we chose to use FOMAML since the core difficulty in CLP ASR is not just parameter-efficient adaptation, but robust generalization to heterogeneous severity conditions. Traditional fine-tuning of the model on the pooled distribution during training is prone to catastrophic fine-tuning/forgetting in the case of limited or distribution-shifted pathological speech samples; catastrophic forgetting in deep learning models is widely studied in continual learning community~\cite{forgetting}. Whisper is a large pre-trained ASR encoder-decoder architecture, which was trained using large-scale weak supervision. This makes the preservation of pre-training knowledge important in the process of fine-tuning. Adapter-based methods and LoRA reduce the number of trainable parameters since most of the pre-trained weights remain constant and there are additional trainable modules and/or low-rank weight matrices added to the model; yet, they do not explicitly model support-query adaptation~\cite{peft}. Knowledge distillation may be helpful for model compression or knowledge transfer from the teacher network to the student; however, the success of knowledge distillation is strongly dependent on the accuracy of the teacher predictions~\cite{distillation}. 
Reason for selecting meta-learning is that CLP ASR needs to have good ability for adapting to varying conditions of severity. Selection of FOMAML is due to the fact that the support-query architecture enables the learning of adaptable parameters of the model from a few support conditions and generalizing them for the query conditions for CLP speech. As compared to MAML, FOMAML does not involve computation of second-order gradients.
In contrast, the FOMAML algorithm is developed on the basis of MAML, in which model parameters are tuned in such a way that a few gradient steps on support data give rise to good performance of the algorithm on queries; the first order approaches have the same adaptation objective without the need to compute second order derivatives~\cite{MAML,FOMAML1}.

Model-agnostic meta learning (MAML) was first introduced as a meta-learning approach applicable across different models to identify initialization parameters that support fast adaptation with only a few gradient updates~\cite{MAML}. One limitation of standard MAML is that full MAML includes higher-order gradient terms during optimization which becomes costly when applied to high-parameter models. First-order MAML provides a more efficient approximation by approximating the meta-gradient using only first-order terms without changing the main goal of obtaining rapidly adaptable parameters~\cite{FOMAML1}. Since small changes in hyperparameters can influence MAML performance and may not always converge smoothly later variants incorporated training refinements to support smoother convergence and enhance performance on unseen tasks~\cite{antoniou2019train}.
Meta-learning approaches have also been explored for sequence modeling and speech recognition under limited-data conditions~\cite{hsu2019meta}. In ASR settings where only a small amount of speech data is available MAML learns model parameters that can be quickly adjusted for a new speech domain with accented target-language data~\cite{winata2020crossaccent}.
A similar idea has been explored in low-resource neural machine translation~\cite{gu2018meta}. This line of work motivates the use of FOMAML for adapting ASR models to disordered speech where clinical severity introduces heterogeneous acoustic and articulatory patterns resulting in a challenging adaptation setting with scarce data. The relationship between meta-learning and personalized federated learning also indicates a possible future direction where clinical centers, speakers and severity groups can represent heterogeneous federated clients~\cite{fallah2020personalized}.
Proposed by Nichol et al. (2018)~\cite{FOMAML1}, FOMAML simplifies the meta-update by ignoring second-order derivatives. It approximates the meta-gradient as the gradient of the query loss at the adapted parameters, treating the adaptation mapping as identity for derivative purposes.
Given these facts, we have explored FOMAML for CLP speech ASR. In a nutshell our contributions are as follows:
\begin{itemize}
    \item We proposed and implemented NA-FOMAML for CLP speech ASR and showcased the enhancements in ASR obtained by keeping normal in inner loop and systemic analysis by placing other severities in the outer loop.
    \item We performed further analysis by fine-tuning different encoder layers.
    \item We reported the phoneme analysis for the same and highlighted which specific phonemes failed in recognition for all the cases.
\end{itemize}

\section{Datasets available}\label{dataset}
There are very few datasets on CLP speech. In this section, we explore an English and a Kannada CLP datasets on which the experiments were carried on.

\subsection{NMCPC-CLP dataset}
The NMCPC-CLP dataset was obtained from the New Mexico Cleft Palate centre by Anil et. al ~\cite{jawed}. This is a private dataset spoken in English language. It comprises 41 CLP speakers out of which 22 are male speakers and 19 are females speakers. Also, a set of 24 normal speakers speaking similar utterances is present. A total of 76 sentences are spoken, each speaker speaking a subset of these total utterances. The CLP speakers were categorized into mild, moderate and severe. The age group of the CLP speakers is 9.2 ± 3.3 years.
\subsection{CLP speech data from AIISH}
This dataset was developed by the students of IIT Guwahati from the All India Institute of Speech and Hearing. A total of 42 speakers recorded their speech with 18 CLP female speakers and 24 male CLP speakers. Speech of 22 Normal female Kannada speakers and 20 male Kannada speakers with parallel syntax is also present. The participants were native Kannada speakers who either had a repaired cleft lip palate or a repaired cleft palate with age between 7-12 years. They did not have any other congenital syndromes or hearing impairment. Speech of children with normal speech speaking the same utterances is also present.

Both the AIISH and NMCPC datasets were partitioned such that the speakers in the training and development sets were disjoint from those in the evaluation set. Thus, the external evaluation set contained only unseen speakers.

To ensure that the observed performance differences were due to the composition of the training data, rather than differences in the number of training utterances, a controlled sampling strategy was adopted for both datasets. For each dataset, the total number of training files in every configuration was fixed to the number of available normal-speech training files. The selected samples were then distributed as evenly as possible across the severity groups included in each configuration.

For the NMCPC dataset, the available training pool consisted of 280 normal, 246 mild, 204 moderate, and 199 severe speech files. Since the normal-speech partition contained 280 files, the total training size was fixed at 280 files for all controlled configurations. The Normal-only configuration used all 280 normal files. The Normal+Mild configuration included 140 normal and 140 mild files. The Normal+Mild+Moderate configuration included 93 normal, 93 mild, and 94 moderate files, while the Normal+Mild+Moderate+Severe configuration included 70 files from each severity group. The CLP-only reference configuration was constructed using 93 mild, 93 moderate, and 94 severe files, again yielding a total of 280 training files. In addition, 70 development files were merged from the corresponding partitions. The evaluation set contained 264 files, with 66 files each from the normal, mild, moderate, and severe groups.

For the AIISH dataset, the available training pool consisted of 304 normal, 302 mild, 247 moderate, and 76 severe speech files. Therefore, the total number of training files was fixed at 304 files for all controlled configurations. The Normal-only configuration used all 304 normal files. The Normal+Mild configuration included 152 normal and 152 mild files. The Normal+Mild+Moderate configuration included 101 normal, 101 mild, and 102 moderate files. For the Normal+Mild+Moderate+Severe configuration, 76 files were selected from each group, resulting in 304 training files. The CLP-only reference configuration consisted of 101 mild, 101 moderate, and 102 severe files, also totaling 304 files. In addition, 76 development files were merged from the relevant partitions. The evaluation set contained 152 files, with 38 files each from the normal, mild, moderate, and severe groups.

This controlled design ensured that all data-composition configurations within a dataset used the same total number of training files, while only the severity composition and diversity were varied. Therefore, the resulting comparisons more directly reflect the effect of severity-aware data inclusion on ASR accuracy and fairness, rather than being influenced by differences in training-set size.

\section{Model-Agnostic Meta-Learning and FOMAML}\label{FOMAML}
Conventional supervised learning trains a model for a fixed task setting within a fixed data domain. Conversely meta-learning  trains a model across a distribution of tasks such that using only a small number of examples or gradient steps it can rapidly adapt to a new task. This is specifically advantageous for the tasks involving heterogeneous data distributions such as pathological speech recognition, where speech characteristics vary substantially across various severity levels such as normal, mild, moderate and severe CLP speech.

Model-Agnostic Meta-Learning (MAML) aims to learn an initial set of model parameters $\phi$ that can be adapted to a given task in a short span of time. For each training episode, the data is divided into a support set $\mathcal{D}_{\text{S}}$ and a query set $\mathcal{D}_{\text{Q}}$. The adaptation set is used to perform short-term adaptation to a given task while the query set estimates the generalization ability of the task-adapted model. FOMAML makes optimization computationally cheap by ignoring second-order derivative terms. It uses the gradient of the query loss with respect to the adapted parameters to approximate the meta-gradient. The first-order meta-update can be written as denoted in equation 1.

For one gradient update, the adapted parameters $\phi'$ are computed as
\begin{equation}
\phi' = \phi - {\eta_{in}} \nabla_{\phi} \mathcal{L}_{\text{S}}(\phi)
\end{equation}

\begin{equation}
\phi \leftarrow \phi - {\eta_{out}} \nabla_{\phi} \mathcal{L}_{\text{Q}}(\phi')
\end{equation}
where $\phi$ in equation 1 represents an initial set of model parameters which are the base model parameters of whisper-small model and $\phi'$ denotes the set of adapted model parameters. The learning rate of the inner and outer-loop is given by ${\eta_{in}}$ and ${\eta_{out}}$ respectively. The support set and query-set losses are denoted as $\mathcal{L}_{\text{S}}(\phi)$ and $\mathcal{L}_{\text{Q}}(\phi')$ respectively. This equation is the same for both FOMAML and MAML. In equation 2, it can be seen that the updated parameters along with the initial parameters now contribute to the new parameters denoted by $\phi$ again.
The same equation is used for MAML except that the gradient is computed through $\phi'$, hence would have $\nabla_{\phi'}$ in place of $\nabla_{\phi}$. Since $\phi'$ itself depends on $\phi$, the chain rule introduces second-order terms. As a result of which MAML gets computationally expensive for large encoder-decoder architectures such as Whisper.
The formulation of FOMAML encourages the model to learn an initialization that can be rapidly adapted to new or shifted speech conditions. 

\begin{figure}
\centering
\includegraphics[height= 200pt,width=250pt]{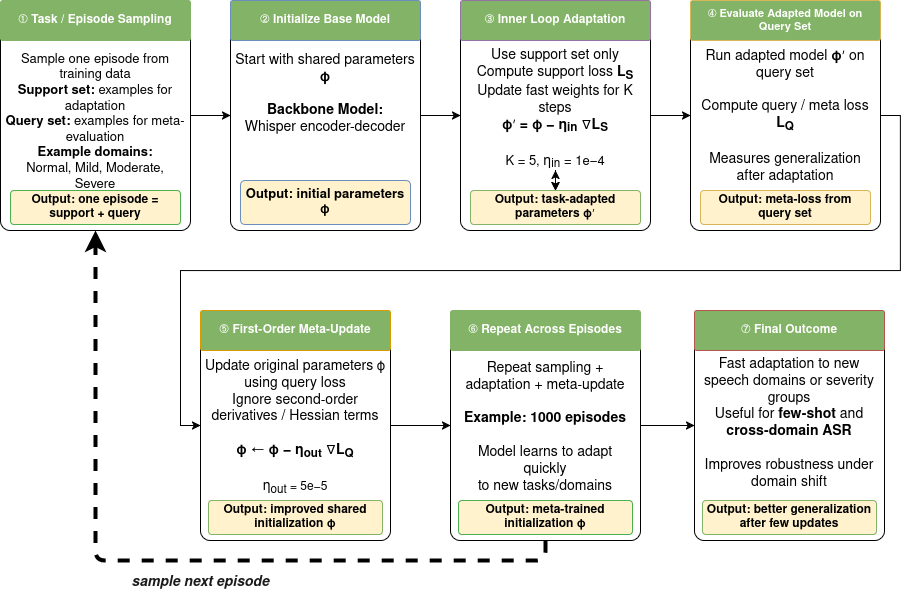}
\caption{Block diagram of FOMAML}
\label{fig4}
\end{figure}

This approximation is highly advantageous, as it reduces computational cost and memory usage. It preserves the key meta-learning objective of learning parameters that are easy to adapt in applications which involve large ASR models. FOMAML therefore provides a favorable trade-off between computational feasibility and adaptation ability for Whisper-based CLP speech recognition. Contrary to standard full fine-tuning, where the model parameters are directly optimized on the available training data as shown in the equation: 
\begin{equation}
\phi \leftarrow \phi - \eta \nabla_{\phi} \mathcal{L}_{\text{train}}(\phi)
\end{equation}
where $\eta$ is the learning rate and $\mathcal{L}_{\text{train}}(\phi)$ is the training loss. So, instead of the existing approach of equation 3 or the second order approach of MAML, FOMAML achieves this feat of better performance by introducing an explicit support-query structure during training.

Although the conventional fine-tuning approach is simple, there are chances of overfitting to the dominant training distribution. As a consequence, it may not generalize well across unseen severity conditions. This is a major concern when it comes to CLP speech recognition, because the severity categories differ substantially in resonance, articulation and intelligibility. This limitation is overcomed by FOMAML since it optimizes the model not only for direct training performance, but also for post-adaptation performance on a separate query set, thereby making it suitable for learning robust cross-severity representations.

\section{Proposed Normal-Anchored First-Order Model-Agnostic Meta-Learning  Approach (NA-FOMAML) for CLP Speech Recognition}\label{NA-FOMAML}

CLP speech due to high variability in its severity levels exhibits a strong domain shift as various factors alter its characteristics such as 
hypernasality, articulation distortion, resonance imbalance and compensatory articulation errors. Due to this difference models are not able to generalize across all severity types as mostly the models are trained only on one severity group. In a similar manner, direct finetuning on mixed pathological data has a higher chance of becoming unstable or biased toward the dominant or easier group. To tackle this problem on a conceptual level this paper proposes a Normal-Anchored FOMAML strategy based approach, where normal speech is used in the support set in the inner loop adaptation, while the outer loop introduces controlled pathological diversity.

\begin{figure}
	\centering
	\includegraphics[height= 280pt,width=250pt]{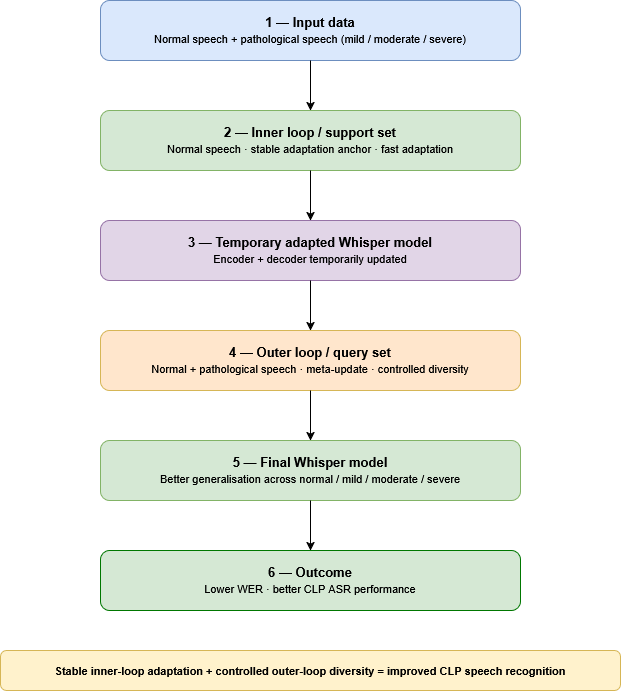}
	\caption{Block diagram of proposed NA-FOMAML}
	\label{fig4}

\end{figure}

The reason behind using Normal speech as a stable adaptation anchor is because it provides relatively stable acoustic and linguistic structure, standard resonance, clearer phoneme realization and higher intelligibility. Let the support set be denoted as

\begin{equation}
\mathcal{D}_{\text{S}} = \mathcal{D}_{\text{no}}
\end{equation}

Using this support set, the model performs a temporary adaptation step:

\begin{equation}
\phi' = \phi - {\eta_{in}} \nabla_{\phi} \mathcal{L}_{\text{norm}}(\phi)
\end{equation}

where $\mathcal{L}_{\text{norm}}(\phi)$ is the loss computed on normal speech. This step allows the model to adapt from a stable and clean and intelligible speech condition rather than being immediately exposed to highly distorted pathological speech during the inner update.

For the full normal-to-all-severity setting, the query set can be written as
\begin{equation}
\mathcal{D}_{\text{Q}} =
\mathcal{D}_{\text{no}}
\cup
\mathcal{D}_{\text{mi}}
\cup
\mathcal{D}_{\text{mo}}
\cup
\mathcal{D}_{\text{se}}
\end{equation}

The outer-loop objective is therefore to first update the base model parameters such that the adapted model performs well across multiple severity groups:

\begin{equation}
\phi \leftarrow \phi - {\eta_{out}} \nabla_{\phi'} 
\mathcal{L}_{\text{Q}}(\phi')
\end{equation}

The interpretation of the expected sampling distribution of query loss may be expressed as a severity-combined objective:

\begin{equation}
\mathcal{L}_{\text{Q}} =
\pi_{\text{no}}\mathcal{L}_{\text{no}}
+
\pi_{\text{mi}}\mathcal{L}_{\text{mi}}
+
\pi_{\text{mo}}\mathcal{L}_{\text{mo}}
+
\pi_{\text{se}}\mathcal{L}_{\text{se}}
\end{equation}

where $\pi_{\text{no}}$, $\pi_{\text{mi}}$, $\pi_{\text{mo}}$ and $\pi_{\text{se}}$ denote the relative contributions of each severity group or in other words the sampling probabilities of the groups. In this setup, the query set composition is controlled by the selected outer-loop configuration from the set: No $\rightarrow$ No, No $\rightarrow$ NoMi, No $\rightarrow$ NoMiMo, or No $\rightarrow$ NoMiMoSe.
Because the utterances of queries are selected uniformly across the pooled outer-loop dataset without severity balancing, the expected probability of the utterances in severity level was: \[
\pi_Q = \frac{n_Q}{\sum_{h \in \mathcal{Q}} n_h}
\] where, say if \[
\mathcal{Q} = \{\text{no}, \text{mi}, \text{mo}\}.
\]
These probabilities reflect the actual empirical distribution of the samples selected given by: \[
\pi_Q =
\begin{cases}
\dfrac{n_Q}
{n_{\mathrm{no}} + n_{\mathrm{mi}} + n_{\mathrm{mo}}},
& Q \in \{\mathrm{no}, \mathrm{mi}, \mathrm{mo}\}, \\[8pt]
0, & Q = \mathrm{se}.
\end{cases}
\]
where $n_{\text{no}}$, $n_{\text{mi}}$, $n_{\text{mo}}$ and $n_{\text{se}}$ denote the number of normal, mild and moderate utterances respectively.
The main motivation for this Normal-Anchored design is that the inner loop should remain stable during meta learning, while the outer loop should expose the model to increasing levels of pathological variability. The normal anchor prevents unstable adaptation during the inner loop by using Normal speech as the stable reference and the outer-loop diversity encourages the model to generalize beyond normal speech. Thus, the model learns an initialization that can adapt to normal speech while maintaining robustness at all levels of severity of CLP. However, the normal speech data was not confined to only inner loop rather it was retained in the outer loop as well. The reason being that the model should not only adapt from normal speech, but also retain normal speech recognition while learning to generalize toward pathological severity groups. The final evaluation is performed separately for normal, mild, moderate and severe groups using WER. As pathological speech performance is equally important as the performance of the normal group, a CLP macro WER can be used to summarize pathological recognition performance:

\begin{equation}
\text{WER}_{\text{CLP-macro}} =
\frac{
\text{WER}_{\text{mi}} +
\text{WER}_{\text{mo}} +
\text{WER}_{\text{se}}
}{3}
\end{equation}

This metric gives equal importance to mild, moderate and severe speech, preventing the overall score from being dominated by normal speech. The overall WER can also be computed over all utterances as:

\begin{equation}
\text{WER}_{\text{overall}} =
\frac{S + D + I}{N} \times 100
\end{equation}

where $S$, $D$ and $I$ represent the number of substitutions, deletions and insertions, respectively and $N$ is the number of reference words.
The performance metrics include the pooled WER which was calculated to weigh the normal speech and CLP speech equally. The pooled WER, ${\mathrm{WER}_\mathrm{Pooled}}$ is calculated as:


\begin{align}
\mathrm{WER}_{\mathrm{Pooled}}
=
\frac{\sum S + \sum D + \sum I}{\sum N}
\times 100
\end{align}
where, ${\sum S}$=Number of substituted words, ${\sum D}$=Number of deleted words, ${\sum I}$=Number of insetrted words, ${\sum N}$=Number of reference words.

The Normal-Anchored FOMAML framework therefore preserves the reliable structure of normal speech while explicitly optimizing the model for cross-severity generalization by combining stable inner-loop adaptation with controlled outer-loop pathological diversity. This design is suitable for CLP speech recognition because it allows Whisper to adapt to limited and highly variable CLP speech more effectively than direct fine-tuning.

\subsection{Fairness as a metric}\label{ob_study2}

Inspired by Howard et al.~\cite{fairness1} and Liang et al.~\cite{fairness2}, we evaluate the fairness of the ASR system using fairness-oriented performance metrics. The Fairness Score (FS) is formulated as a weighted measure that jointly accounts for overall recognition accuracy and performance imbalance across groups. Specifically, it combines the negative average error rate with the error disparity between two groups. Here, the average error rate is computed as the mean error rate across the two groups, whereas the error disparity is measured as the absolute difference between their error rates. The Fairness Score (FS) is defined as follows:


\begin{equation}
FS = -\alpha \cdot \text{Average Error Rate} - \beta \cdot \text{Error Disparity}, \quad \alpha, \beta \geq 0
\label{eq:fairness_score}
\end{equation}

where:

1. \textbf{Average Error Rate:}

\begin{equation}
\text{Average error rate} = \frac{\text{error}(G_1) + \text{error}(G_2)}{2}
\label{eq:aer}
\end{equation}


Here, \(\text{error}(G_1)\) and \(\text{error}(G_2)\) are the error rates for the two groups, here the group $G_1$ indicates Normal and $G_2$ refers to CLP.

2. \textbf{Error disparity:}

\begin{equation}
\text{Error Disparity} = \left| \mathrm{error}(G_1) - \mathrm{error}(G_2) \right|
\label{eq:ed}
\end{equation}


3. \(\alpha\) and \(\beta\) are the coefficients to balance the importance of overall error minimization and fairness (disparity minimization).

 The range of $FS$ is $ -\infty\leq \text{FS} \leq 0$. The value of \textbf{\emph{$FS$ closer to zero}} signifies \textbf{\emph{better fairness}} of the system, and vice versa. That is, the value of $FS$ represents the \textbf{\emph{degree of fairness}} of the system.

\section{Experimental Setup}\label{exp}

The main objective of this work was to adapt the Whisper base model for CLP speech recognition. The dataset partitioning was done in an 80:20 format for train and test. Again the train was partitioned in the same format for train and development sets. Each set contained 4 partitions normal, mild, moderate and severe. The meta training was performed on train set and the best checkpoint was selected from the Development set after applying early stopping. The experiments were conducted in English and Kannada languages. The Kannada language transcripts were written in Roman transliterated form. Hence, in the decoding stage the language settings were kept as English.

The proposed training technique follows the FOMAML strategy. Each episode comprises an inner-loop adaptation followed by an outer-loop update using support and query set, respectively. Given the severe speech deformities in CLP speech categories and the difference in their distributions, the design of the datasets in each of inner and outer loops required careful consideration. The support set contained normal speech and the query set had normal along with CLP speech severities added from time to time. This approach was followed to check for the best support and query set design so that the model  generalizes towards pathological speech by adapting from both normal intelligible speech. 
Owing to the less data size, models with fewer parameters were taken into consideration. OpenAI's whisper-small model is a careful consideration for base ASR model, given the parameters of the initial layers were kept frozen. Also, since overfitting is a real issue in fine-tuning Whisper models and only training the last few layers do not provide proper learning to the models, a detailed experiment was carried out once by freezing encoder and then by unfreezing encoder layers in sets. It included unfreezing encoder layers 0--11, 0--5, 4--11, 6--11, 8--11, the decoder and the projection head. The encoder convolution layers were kept frozen to keep the lower-level acoustic representations intact, thus allowing the mid and final encoder layers and decoder to adapt to CLP speech characteristics.
The model training was carried out on a total of 300 episodes per fold with an inner learning rate of $1 \times 10^{-5}$, an outer learning rate of $5 \times 10^{-6}$. The support and query batch of 8 each used 1 inner adaptation step with regularisation, which included label smoothing = 0.05 and gradient clipping = 1.0 along with weight decay = 0.01. The evaluation on the development set was evaluated every 25 episodes and early stopping using patience = 3.

The best model was selected based on outer CLP macro WER instead of overall WER to avoid model selection bias from just the normal group as shown in equation 9. This CLP macro WER was used instead of overall WER to prevent the normal group from biasing the model selection. The models from the 5-folds were later averaged in a weighted manner to reduce the variance. The inferencing was done on this final robust model. As models are susceptible to overfitting due to the unevenness in data distribution, a safety mechanism that ensures loss below \text{loss floor = 0.01}, \text{loss floor patience = 2}, \text{loss floor metric = query} and \text{loss stop min episode = 125} was introduced. Greedy decoding with repetition prevention was adopted during training and evaluation with the following parameters: \text{no repeat ngram size = 2} and \text{length penalty = 0.8}. Evaluation metrics include \text{WER normal}, \text{WER mild}, \text{WER moderate}, \text{WER severe}, \text{WER overall}, \text{WER clp macro} and \text{WER macro all groups}.
The ensemble decoding was performed from all the 5 folds. Ensemble decoding retains individual fold models and aggregates the predictions made by them. Hence, in case of any weak or strange predictions by one of the models, it can be corrected by consensus among other models. It is especially important for CLP speech as speaker and severity variability is generally high. 
Finally, a detailed phoneme analysis of hypothesis outputs was performed to analyse the error patterns of all the severities.

\section{Results and Discussion}\label{res}

\begin{table*}[t]
\centering
\caption{Comparison of pretrained Whisper, conventional full encoder-decoder fine-tuning, and FOMAML full encoder-decoder training strategies for NMCPC dataset. All systems are evaluated on the same fixed evaluation set. WER values are reported in percentage. All Macro is the unweighted average across normal, mild, moderate and severe groups, while CLP Macro is the unweighted average across mild, moderate and severe groups.}
\label{tab:whisper_fomaml_comparison}
\resizebox{\textwidth}{!}{
\begin{tabular}{llcccccccc}
\hline
Method & Training configuration
& Normal & Mild & Moderate & Severe & All Macro & CLP Macro & Pooled WER & FS \\
\hline

Pretrained Whisper
& No fine-tuning
& 16.80 & 32.41 & 58.66 & 52.55 & 53 & 65.07 & 104.13 & -44.15 \\

Conventional fine-tuning
& Normal only
& 5.20 & 15.02 & 42.52 & 35.74 & 36.14 & 46.45 & 81.82 & -33.14 \\

CLP-anchored FOMAML
& Inner: CLP; Outer: Normal
& 9.20 & 26.88 & 55.91 & 45.15 & 45.52 & 57.62 & 90.08 & -40.55 \\

NA-FOMAML
& Inner: Normal; Outer: Normal
& 5.60 & 16.60 & 44.49 & 36.84 & 37.23 & 47.77 & 82.23 & -34.04 \\

NA-FOMAML
& Inner: Normal; Outer: No+Mi
& 4 & 8.3 & 22.44 & 24.02 & 24.39 & 31.18 & 62.81 & -22.02 \\

NA-FOMAML
& Inner: Normal; Outer: No+Mi+Mo
& 4.40 & 5.53 & 16.14 & 52.07 & \textbf{19.54} & \textbf{24.58} & \textbf{19.22} & \textbf{-17.02} \\
NA-FOMAML
& Inner: Normal; Outer: No+Mi+Mo+Se
& 6 & 7.11 & 18.90 & 21.42 & 21.74 & 26.99 & 54.96 & -18.42 \\
\hline
\end{tabular}
}
\end{table*}

\begin{table*}[t]
\centering
\caption{Comparison of pretrained Whisper, conventional full encoder-decoder fine-tuning, and FOMAML full encoder-decoder training strategies for AIISH dataset. All systems are evaluated on the same fixed evaluation set. WER values are reported in percentage. All Macro is the unweighted average across normal, mild, moderate and severe groups, while CLP Macro is the unweighted average across mild, moderate and severe groups.}
\label{tab:whisper_fomaml_comparison_aiish}
\resizebox{\textwidth}{!}{
\begin{tabular}{llccccccccc}
\hline
Method & Training configuration
& Normal & Mild & Moderate & Severe & All Macro & CLP Macro & Pooled WER & FS \\
\hline

Pretrained Whisper
& No fine-tuning
& 108.26 & 162.39 & 164.46 & 175.83 & 152.74 & 167.56 & 152.61 & -98.48 \\

Conventional fine-tuning
& Normal only
& 3.31 & 66.67 & 52.07 & 108.33 & 57.59 & 75.69 & 57.41 & -55.75 \\

CLP-anchored FOMAML
& Inner: CLP; Outer: Normal
& 3.31 & 75.21 & 63.64 & 111.67 & 63.46 & 83.51 & 63.26 & -61.60 \\

NA-FOMAML
& Inner: Normal; Outer: Normal
& 2.48 & 63.25 & 51.24 & 110.83 & 56.95 & 75.11 & 56.78 & -55.54 \\

NA-FOMAML
& Inner: Normal; Outer: No+Mi
& 3.31 & 21.37 & 28.10 & 88.33 & 35.28 & 45.93 & 35.28 & -33.62 \\

NA-FOMAML
& Inner: Normal; Outer: No+Mi+Mo
& 2.48 & 23.08 & 19.83 & 66.67 & 28.01 & 36.53 & 27.97 & -26.73 \\
NA-FOMAML
& Inner: Normal; Outer: No+Mi+Mo+Se
& 2.48 & 19.66 & 14.05 & 57.50 & \textbf{23.42} & \textbf{30.40} & \textbf{23.38} & \textbf{-22.14} \\
\hline
\end{tabular}
}
\end{table*}

\begin{table*}[t]
\centering
\caption{Comparison of encoder layer fine-tuning strategies for the NoMiMo configuration on the NMCPC and NoMiMoSe configuration on AIISH dataset. WER values are reported in percentage.}
\label{tab:nomimo_encoder_comparison_both}
\footnotesize
\setlength{\tabcolsep}{4pt}
\renewcommand{\arraystretch}{0.95}

\begin{tabular}{llcccccccc}
\hline
Dataset & Encoder setting & Normal & Mild & Moderate & Severe & All Macro & CLP Macro & Pooled WER & FS \\
\hline
\multirow{6}{*}{NMCPC}
& Frozen encoder & 5.60 & 9.88 & 31.89 & 69.01  & 29.09 & 36.93 & 28.73 & -25.93 \\
& Enc. 8--11     & 5.60 & 6.32 & 32.68 & 61.57 & 26.54 & 33.52 & 26.23 & -23.43 \\
& Enc. 6--11     & 4.00 & 6.32 & 23.23 & 57.02  & 22.64 & 28.86 & 22.32 & -20.32 \\
& Enc. 4--11     & 4.40 & 5.93 & 19.29 & 54.55 & 21.04 & 26.59  & 20.72 & -18.52 \\
& Enc. 0--11     & 4.40 & 5.53 & 16.14 & 52.07 & \textbf{19.54} & \textbf{24.58}  & \textbf{19.22} & \textbf{-17.02} \\
& Enc. 0--5 & 4.80 & 6.32 & 24.80 & 55.79 & 22.93 & 28.97  & 22.93 & -25.30 \\
\hline
\multirow{6}{*}{AIISH}
& Frozen encoder & 2.48 & 23.93 & 27.27 & 58.33 & 28 & 36.51 & 27.97 & -26.73  \\
& Enc. 8--11     & 2.48 & 25.64 & 23.14 & 60 & 27.82 & 36.26 & 27.77 & -26.53  \\
& Enc. 6--11     & 2.48 & 23.08 & 26.45 & 63.33  & 28.83 & 37.62 & 28.81 & -27.57  \\
& Enc. 4--11     & 2.48 & 17.95 & 19.83 & 65 & {26.32} & {34.26}  & 26.30 & -25.06 \\
& Enc. 0--11     & 2.48 & 19.66 & 14.05 & 57.50  & \textbf{23.42} & \textbf{30.40}  & \textbf{23.38} & \textbf{-22.14} \\
& Enc. 0--5      & 3.31 & 23.08 & 19.01 & 62.50 & 26.97 & 34.86  & 26.93 & -25.27 \\
\hline
\end{tabular}
\end{table*}

\subsection{Phoneme-Level Error Analysis}

Error analysis at the phoneme level was carried out to understand how different configurations of outer loop severity impact the errors made for different broad phonemes. In order to carry out the analysis, the word-level reference hypothesis alignment was used along with an orthographic-based phoneme category mapping. 
For NMCPC and AIISH, using direct pretrained Whisper provided CLP macro WER of $65.07\%$ and $100\%$. Followed by conventional fine-tuning CLP macro of  $46.45\%$ and $75.69\%$ respectively. In FOMAML, when pathological data is provided in the inner loop followed by typical speech data in the outer loop, the CLP macro WER deteriorates to $57.62\%$ and $83.51\%$ respectively. Henceforth the selection of Normal is made for the inner loop. It appears that the No $\rightarrow$ No showed significant improvement in WER for both datasets $47.77\%$ and $75.11\%$ as compared to the previous combinations, but the outer-loop severity configuration was poorly generalized for pathological speech recognition. So, consequently experiments were performed keeping normal as anchored data and gradually incorporating pathological speech to the outer loop according to severity. Although the configuration was able to preserve WER low for normal speech, its performance decreased significantly for mild, moderate and severe groups. The poor generalization suggests that training only on normal speech data is likely to cause the system to learn only the acoustic and phonological properties of the normal utterances while neglecting the variability of cleft-related articulations.

For NMCPC, introducing pathological speech into the outer-loop query set resulted in improvement across different severities. With the use of encoder fine-tuning of all 0--11 layers, improved the CLP macro WER to $24.58\%$ followed by 
layers 6--11 for fine-tuning, the optimal configuration of No $\rightarrow$ NoMiMo has WERs of 4.00\%, 6.32\%, 23.23\% and 57.02\% for normal, mild, moderate and severe speech, correspondingly. Similar trends were also observed for encoder layers 4--11, in which No $\rightarrow$ NoMiMo produced the best fine-tuning results, with WER scores of 4.40\%, 5.93\%, 19.29\% and 54.55\% in the respective severity groups. The layers 8--11 and 0--5 however gave the least improvement with CLP Macro being $33.52\%$ and $28.97\%$, but still better than Frozen encoder which is $36.93\%$.

From this result, it can be inferred that mild and moderate speech in the outer loop can help capture intermediate pathological features of severe conditions during adaptation. However, using severe speech in addition to these categories (i.e., No $\rightarrow$ NoMiMoSe) for NMCPC yielded poorer results for all severity levels, potentially because of the increased variance introduced by such samples within the current training framework.
The phoneme-level analysis of NMCPC provides evidence for this finding. The outer loop training on the normal only subset produced extensive category error rates among all phoneme classes, specifically affricates, nasals, liquids, fricatives, plosives and vowels. With No $\rightarrow$ NoMiMo adaptation, category errors for mild and moderate samples were significantly decreased, yet severe samples remained problematic. With the same layers' fine-tuning process, severe speech samples continued to show higher rates of category errors, namely affricates, nasals, liquids and pressure consonants, whose phonemes include /\textesh/, /t\textesh/, /d\textyogh/, /\ng/, /j/, /w/, /n/, /m/, /s/, /f/, \textipa{/E/} and \textipa{/I/}.

In the case of AIISH, fine-tuning of encoder 0--11 improved the CLP macro WER to $30.40\%$ followed by fine-tuning of encoder 6--11 and 4--11. Both yielded benefits from having the full severity set of samples in the outer loop. In encoder layers 6--11, No $\rightarrow$ NoMiMoSe resulted in the best CLP Macro WER of $37.62\%$. Meanwhile, when using encoder layers 4--11, the aforementioned configuration also produced the best CLP Macro WER, at 34.26\% and All Macro WER, at 26.32\%. At this configuration, the encoder layers 4--11 yielded WERs of 2.48\%, 17.95\%, 19.83\% and 65.00\% for normal, mild, moderate and severe speech, respectively. 
These results indicate that having severe AIISH samples did not negatively affect outer loop optimization and may actually contain useful information for adaptation. Results of the AIISH phoneme-level study confirm the notion that even under the optimal configuration, severe speech remains the hardest type. At encoder layers 4--11 using the No $\rightarrow$ NoMiMoSe configuration, severe speech generated extremely high category errors for affricates, nasals, vowels, fricatives, plosives and liquids. 
The most vulnerable phones were \textipa{/n/}, \textipa{/S/}, \textipa{/E/}, 
\textipa{/g/}, \textipa{/dZ/}, \textipa{/r/}, \textipa{/l/}, 
\textipa{/I/}, \textipa{/d/} and \textipa{/ae/}. 
The highest absolute error counts were observed for \textipa{/ae/}, 
\textipa{/t/}, \textipa{/d/}, \textipa{/I/} and \textipa{/p/}. The category failures mostly include nasals, affricates, liquids, vowels, and plosives.
The high errors of plosives, fricatives, and affricates are due to the challenges in the oral pressure and release cues. But the high errors of nasals and liquids indicate that resonance and transition problems are equally important for ASR.
In summary, this analysis proves that the severity-aware normal-anchored FOMAML approach increases not only WER but also the robustness of Whisper ASR.
In summary, tuning all encoder layers was more effective at balancing accuracy and efficiency than tuning partial encoder layers in both datasets followed by 4–11 and 6-11. For this dataset as well, the layers 8--11 and 0--5 gave the least improvement with CLP Macro being $36.26\%$ and $34.86\%$, but still better than Frozen encoder which is $36.51\%$.
These findings support our hypothesis that mid-to-upper layer fine-tuning outperforms only top-layer fine-tuning and full finetuning gave the best WER performance without incurring additional training costs and risks of overfitting.
From this comparison, we conclude that CLP speech changes resonance, nasality, burst clarity, frication noise and vowel quality which are captured by lower layers (0-3), followed by distorting pressure consonants, fricatives, affricates, nasals, liquids, substitutions captured by middle layers (4-7) and finally the upper layers (8-11) which alter Higher-level speech-unit and context representation such as token-aligned acoustic patterns, word-level cues, language-dependent abstraction. Hence fine-tuning of lower layers improves the overall performance. Also, including severity-aware sampling into the outer loop contributes to the optimization of CLP ASR models. NMCPC yielded the best results by applying severity-aware sampling within the outer loop that includes all levels (normal, mild and moderate). AIISH, conversely, showed the best performance through severity-aware sampling applied to all four outer loop levels (normal, mild, moderate and severe). Thus, severity-aware meta-training should account for the unique characteristics of every specific dataset. Still, severe speech appears to be the main source of ASR errors in both of them in cases of affricates, fricatives, nasals, liquids, vowels and pressure consonants.

\subsection{Key Observations from Phoneme Analysis}

Regarding the full-encoder FOMAML scenario on the NMCPC dataset, the phoneme-level analysis showed that increasing the outer-loop severity diversity from the normal-only condition to the NoMiMo condition substantially reduced CLP phoneme error rates. The most notable reductions were observed for plosives (/p/, /b/, /k/, /ks/), fricatives (/v/, /z/), affricates (/d\textyogh/), liquids (/w/, /r/), and vowels (/u\textlengthmark/, /a\textupsilon/, /e\textsci/). However, further extending the outer loop by including severe speech did not provide additional improvement. In fact, the NoMiMoSe configuration resulted in a higher WER than NoMiMo, suggesting that severe NMCPC speech may introduce greater acoustic variability or gradient conflict during meta-training.

For NMCPC, the phoneme-category analysis further indicated that encoder-layer adaptation reduced most CLP phoneme errors compared with the frozen-encoder baseline. The largest improvements were observed for nasals (/m/, /n/), fricatives (/f/, /v/, /s/, /\textesh/), vowels (/\textsci/, /\textepsilon/), and affricates (/t\textesh/). Full encoder adaptation produced the strongest overall reductions for fricatives, vowels, plosives, and affricates. Nevertheless, mid-to-upper encoder-layer adaptation also yielded strong improvements for clinically relevant phoneme categories, particularly nasals and fricatives, while using fewer trainable parameters.

The AIISH full-encoder FOMAML phoneme analysis showed that increasing the severity diversity in the outer loop improved phoneme-level robustness for CLP speech. Compared with the normal-only outer-loop baseline, the NoMiMoSe configuration achieved the largest reduction in phoneme error rates, especially for nasals (/m/, /n/), fricatives (/s/, /h/), plosives (/b/, /p/, /d/, /k/), and vowels (/u\textlengthmark/, /\textscripta/, /\textturnv/, /\textsci/). These findings suggest that including severe CLP speech in the outer loop improves generalization to pathological speech variability. However, the improvement was not uniform across all phonemes, as liquids such as /r/ showed only marginal improvement compared with the NoMiMo configuration.

According to the AIISH NoMiMoSe phoneme-level analysis, full encoder fine-tuning was the most effective in reducing errors across most phoneme categories. Compared with the frozen-encoder configuration, Enc.~0--11 improved the recognition of fricatives such as /s/, /\textesh/, and /v/, plosives such as /b/ and /k/, and vowels such as /\textturnv/ and /\ae/. However, the effectiveness varied across layer configurations. Tuning only the lower encoder layers improved selected isolated phonemes such as /m/ and /i\textlengthmark/, but degraded affricates such as /t\textesh/ and vowels such as /u\textlengthmark/.

Therefore, future work will explore phoneme-aware loss functions, fricative- and affricate-focused data augmentation, severity-aware sampling, and explicit modeling of resonance-related distortions.

\section{Conclusion}\label{Conclusion}

The experiment employed Whisper-small as the base ASR model and adapted it using a first-order bilevel meta-learning strategy. The inner loop used normal speech as the support set, while the outer loop optimized the model on normal, mild, moderate and severe pathological speech samples by sets. A model trained only by normal speech and may not generalize well to mild, moderate, or severe CLP speech as ordinary fine-tuning may learn the dominant/easier patterns. The bilevel strategy, instead of only minimizing training loss directly forces the model to learn parameters that are easy to adapt and useful after adaptation. 
The presented approach learns an initialization that is beneficial for CLP detection through the normal-speech adaptation process. The normal-speech examples in the support set will ensure a consistent inner-loop gradient, whereas the presence of both normal and CLP speech in the query set will guarantee that the adapted parameters are tested in terms of both maintaining the normal-speech performance and being resistant to the pathological one. Therefore, the meta-objective motivates the normal-speech adaptation process to be aligned with CLP detection.
To minimize overfitting, convolutional frontend and the lower encoder layers were frozen while encoder layers 6 to 11 and 4 to 11, the decoder and the output projection layer were fine-tuned. The benefits achieved through encoder-layer fine-tuning are explained by the fact that there was a decrease in representational mismatch between features extracted from pretrained Whisper models and CLP speech acoustics. Lower encoder layers extract local spectral and temporal features, middle layers – phonetic and sub-phonetic features, while the higher layers extract the more contextual speech representations used by the decoder. Due to the fact that CLP speech influences the signal at several levels of analysis such as resonance, pressure consonant production, frication, voicing, and articulatory transitions, it is not sufficient to fine-tune only the higher layers. Fine-tuning increasingly deep encoder layers lets the model reconstruct the distorted acoustic–phonetic representation of CLP speech. It is proven by the highest WER reduction through fine-tuning of Enc. 0–11. The strategies applied to reduce overfitting also included low learning rates, gradient clipping, weight decay, one inner-loop adaptation step and label smoothing. The checkpointing for Development-set was performed according to the minimal outer loop CLP macro WER which varied from normal to normal+mild+moderate+severe. The proposed NA-FOMAML helps the model learn a better adaptation strategy and the performances improved for all severities. Rule-based phoneme-category error statistics and Utterance-level predictions were also generated to support detailed error analysis. However, further improvement on severe partition is required as the WER can still remain high for severe speech since it is very different and underrepresented.

\balance
\bibliographystyle{elsarticle-num} 
\bibliography{cas-refs}
\end{document}

\endinput